\documentclass[preprint,aps,prd,nofootinbib]{revtex4}
\usepackage{amsfonts,amssymb}
\usepackage{mathrsfs}
\usepackage{amsmath}
\usepackage{graphicx}
\usepackage{subfigure}
\usepackage[colorlinks,
            linkcolor=black,
            anchorcolor=blue,
            citecolor=black
            ]{hyperref}

\newcommand{\bea}{\begin{eqnarray}}
\newcommand{\eea}{\end{eqnarray}}
\newcommand{\beq}{\begin{equation}}
\newcommand{\eeq}{\end{equation}}
\newcommand{\nn}{\nonumber}
\def\f{{\mathbf f}}
\def\r{{\mathbf r}}
\def\b{{\mathbf b}}
\def\/{\over}
\makeatletter

\newcommand{\Rmnum}[1]{\expandafter\@slowromancap\romannumeral #1@}
\makeatother

\begin{document}

\title{Minisuperspace quantization of bubbling AdS$_2\times$S$^2$ geometries}
\author{Qinglin Li\footnote{qli20@albany.edu}}
\affiliation{Department of Physics,
University at Albany (SUNY),
Albany, New York, 12222, USA}

\begin{abstract}

We quantize the moduli space of supersymmetric microstates describing four-dimensional black holes with AdS$_2$$\times$S$^2$ asymptotics. To acquire the commutation relations of quantization, we find the symplectic form that is imposed in the Type IIB SUGRA and defined in the space of solutions parametrized by one complex harmonic function in $\mathbf{R}^3$ with sources distributed along closed curves.

\end{abstract}
\maketitle

\baselineskip=16pt

\section{Introduction}

Gauge theory and supergravity are connected by the AdS/CFT correspondence \cite{JM,Gubser,Witten98,Aharony:1999ti}, which allows one to describe supergravity configuration  in terms of microscopic degrees of freedom in field theories on the boundary. The full understanding of the map between the bulk and the gravity sides for AdS$_2$/CFT$_1$ is still missing, although impressive progress has been reported in \cite{Strominger:1998yg,Gibbons:1998fa,Maldacena:1998uz,Blum:1999pc,Michelson:1999zf}.
Recently, regular BPS geometries with AdS$_2$$\times$S$^2$$\times$T$^6$ asymptotics were constructed in \cite{OL,Bianchi}\footnote[1]{Related geometries with flat asymptotics have been discussed in \cite{Pieri}.},
and one can hope to use them to get insight into the structure of the dual field theory. In higher dimensions, where the AdS/CFT duality is well understood, the gravity solutions, known as bubbling geometries \cite{LLM} and fuzzballs \cite{OM,LM02,MF,bena,ks,sd}, can be mapped into the field-theoretic degrees of freedom via quantization of the moduli space \cite{SR,Grant,LM}. In this article we gain some insights into the field theory dual to AdS$_2$ by applying the techniques developed in  \cite{SR,Grant,LM} to the geometries constructed in \cite{OL}.

According to the AdS/CFT dictionary, AdS$_2$$\times$S$^2$ corresponds to the ground state of the dual fields living on the boundary of the AdS space.  The light excitations of the dual fields correspond to strings moving on AdS$_2$$\times$S$^2$, and such strings which turn out to be integrable \cite{Sorokin:2011rr}. Excitations with intermediate energies are mapped into probe D branes whose counterparts in six- and ten-dimensional theories are called giant gravitons~\cite{McGreevy:2000cw,Grisaru:2000zn}. At higher energies the gravitational backreaction of branes must be taken into account, and the corresponding solutions of supergravity were constructed in~\cite{OL}. The higher dimensional counterparts of such gravity solutions have been constructed in \cite{OM,LM02,LLM}, and in \cite{SR,Grant,LM}, the spaces of these supersymmetric geometries were quantized using the method  of Crnkovi$\acute{\text{c}}$-Witten~\cite{witten} and Zukerman~\cite{zukerman}, and a perfect agreement with the expectations from the field theory side \cite{Cor,Ber} was found. Additional applications of this method have also been developed in~\cite{donos,Biswas,boer,boer2,Mandal,Krishnan:2015vha}. In the four-dimensional case we can use the similar quantization to get insights into the dual theory.

The organization of this paper is as follows. In Sec. II we review the regular supersymmetric solutions of supergravity with AdS$_2$$\times$S$^2$$\times$T$^6$ asymptotics, which  are parametrized by two harmonic functions in $\mathbf{R}^3$ with sources distributed along closed curves. In  Sec. III we give the general idea of quantizing geometries from the supergravity action by the approach of Crnkovi$\acute{\text{c}}$-Witten-Zukerman. We apply this method, in the next section, to the bubbling geometries for AdS$_2\times $S$^2$, and the corresponding symplectic form is found. In the last section we give a brief discussion.

\section{BPS solutions of SUGRA}

The simplest way to probe the AdS$_2$/CFT$_1$ duality is to look at strings propagating on
\beq\label{AdSxS}
\text{AdS}_2\times \text{S}^2\times \text{T}^6,
\eeq
and dynamics of such objects was studied in \cite{Sorokin:2011rr,Abbott:2013kka,Hoare:2014kma,Roiban:2014cia}. The heavy excitations are described by spaces which asymptote to (\ref{AdSxS}), and we focus on supersymmetric geometries. Recently, a family of such regular BPS solutions of Type IIB SUGRA was found in \cite{OL}, and the geometry is given by
 \bea\label{Soln}
 ds^2&=&-h^{-2}(dt+V)^2+h^2dx_adx_a+dz^{\dot{a}}d\bar{z}_{\dot{a}},\nn\\
 F_5&=&F\wedge \text{Re}\;\Omega_3-\tilde{F}\wedge \text{Im}\;\Omega_3,~~~\Omega_3=dz_{123},\nn\\
 F&=&-\partial_a A_t(dt+V)\wedge dx^a+h^2\star_3 d\tilde{A}_t,~~~\tilde{A}_t+i A_t={1\/4h}e^{i\beta},\nn\\
 \tilde{F}&=&-\partial_a \tilde{A}_t(dt+V)\wedge dx^a-h^2\star_3 dA_t,~~~dV=-2h^2\star_3d\beta,\nn\\
 \tilde{\eta}&=&h^{-1/2}e^{i\beta\Gamma_5/2}\epsilon,~~~\Gamma^t\Gamma_5\epsilon=\epsilon.\label{GS}
 \eea
Here $a=1,2,3$ and $z_{\dot{a}}=X_{\dot{a}}+iY_{\dot{a}}$ with $\dot{a}=1,2,3$ in which $X_{\dot a}$ and $Y_{\dot{a}}$ are the coordinates for T$^6$. This geometry can be parametrized by two harmonic functions $H_1$ and $H_2$ in $\mathbf{R}^3$,
\bea
H_1&=&h\sin\beta,~~~H_2=h\cos\beta,~~~d\star_3dH_a=0,\nn\\
dV&=&-2\star_3\left[H_2dH_1-H_1dH_2\right],~~~\tilde{A}_t+iA_t={1\/4(H_2-iH_1)}.\label{HF}
\eea
At the points where the two harmonic functions $H_1$ and $H_2$ have sources, the solutions~(\ref{GS}) may be singular. To guarantee the regularity, one has to require that all the sources are distributed along closed curves in $\mathbf{R}^3$, and the harmonic functions
are obtained from:
\bea
 H&=&H_1+iH_2={1\/2 \pi}\int {\sigma\sqrt{(\r-\f)\cdot(\r-\f+\b)}\/(\r-\f)^2}dv+H_{reg},\label{H}\\
&&~~~~~~\b\cdot\dot{\f}=0,~~~~~\b\cdot\b=0, \nn
\eea
where $\f(v)$ is the location of the profile parametrized by $v$ and $\b$ is a complex vector. See \cite{OL} for the detailed discussion.

\section{Symplectic form of Type IIB SUGRA}

A method of quantizing Lagrangian theories was proposed by Crnkovi$\acute{\text{c}}$ and Witten~\cite{witten} and Zukerman~\cite{zukerman}. As a Lagrangian theory, Type IIB SUGRA theory has a symplectic form $\Omega$ defined on the full phase space, which contains the information of the commutation relations. In this paper we will restrict $\Omega$ in the space of solutions. One can write the symplectic form as an integral over the Cauchy surface $\Sigma$:
\beq
\Omega=\int d \Sigma_l J^l.
\eeq
We call $J^l$ a symplectic current, and its explicit form is given by:
\beq
J^l=\sum_i\delta\left[{\partial L\/\partial(\partial_l\phi_i)}\right]\wedge\delta\phi_i,\label{current}
\eeq
where $\phi_i$ runs over all fields in the action. Once the symplectic form is known as
\beq
\Omega={1\/2}\omega^{-1}_{ij}dq_i\wedge dp_j,\label{lsf}
\eeq
one then can encode the Poisson bracket as
\beq
\{q_i,p_j\}_{P.B.}=\omega_{ij}.\label{PB}
\eeq
The commutators are directly obtained from the above brackets by the Dirac prescription.

Since solutions (\ref{Soln}) are supported only by the five-form flux $F_5$, the relevant part of the
Type IIB SUGRA action is given by
\beq
S={1\/2\kappa^2_{10}}\int d^{10}x\sqrt{-g}\left(R-4|F_5|^2\right).
\eeq
The symplectic form does not change as the self-duality constraint $F_5=\star F_5$ is imposed. The symplectic form reads
\beq
\Omega={1\/2\kappa^2_{10}}\int d \Sigma_l (J^l_G+J^l_F),\label{SF}
\eeq
where $J^l_G$ and $J^l_F$ are the gravity and five-form currents, respectively, that can be computed from Eq.~(\ref{current}).

The symplectic current of gravity $J^l_G$ was found by Crnkovi$\acute{\text{c}}$ and Witten,
\beq
J^l_G=-\delta\Gamma^l_{mn}\wedge\delta(\sqrt{-g}g^{mn})+\delta\Gamma^n_{mn}\wedge\delta(\sqrt{-g}g^{lm}).\label{jg}
\eeq
When the metric perturbations take a gauge transformation as
\beq
\delta g_{mn}\rightarrow\delta g_{mn}+\nabla_{(m}\xi_{n)},\label{gauge}
\eeq
$J^l_G$ will change by a total derivative, which keeps the symplectic form~(\ref{SF}) gauge invariant.

Taking the potentials $A_{|k_1\cdots k_4|}$, which satisfy the relation $F_5=dA$, as our basic fields, and applying~(\ref{current}), we obtain $J^l_F$ directly
\bea
J^l_F=-8\delta(\sqrt{-g}F^{l|k_1\cdots k_4|})\wedge \delta A_{|k_1\cdots k_4|},\label{jf}
\eea
where $|i_1i_2i_3i_4|$ means $i_1<i_2<i_3<i_4$.

\section{Minisuperspace Quantization of bubbling geometries for $\mathbf{AdS_2\times S^2}$}

In the parametrization (\ref{Soln}), the AdS$_2$$\times$S$^2$ background is specified by a circular closed curve
\bea
f_1(v)=L\cos{v\/L},~~~f_2(v)=L\sin{v\/L},~~~f_3=0,~~~0\leq v\leq2\pi L,\label{adsf}
\eea
and the related vector $\b$ is given by $\b=2L(\cos {v\/L},\;\sin {v\/L},\;i)$. After
plugging~(\ref{adsf}) and $\b$  into~(\ref{HF}) and~(\ref{H}), it is straightforward to get
\bea
H&=&{L\/\sqrt{r^2+(y-i L)^2}},\nn\\
V&=&-{L\/2}\left[{r^2+y^2+L^2\/\sqrt{4L^2y^2+(r^2+y^2-L^2)^2}}-1\right]d\phi.\label{uh}
\eea
Substituting the above expressions for $H$ and $V$ into~(\ref{GS}), and making the coordinate transformation as
\bea
t={\tilde{t}\/L},~~~
r=L\sqrt{\rho^2+1}\sin\theta,~~~
y=L\rho\cos\theta,~~~
\phi=\tilde{\phi}-t,\label{adsc}
\eea
one  obtains immediately  the geometry of the global AdS$_2$$\times$S$^2$:
\bea
ds^2&=&L^2\left[-(\rho^2+1)d\tilde{t}^2+{d\rho^2\/\rho^2+1}+d\theta^2+\sin^2\theta d\tilde{\phi}^2\right]+dz^ad\bar{z}_a,\nn\\
F_5&=&{L\/4}d\rho d\tilde{t}\wedge\text{Re}(dz_{123})+\text{dual}.\label{ads2}
\eea

The circular profile~(\ref{adsf}), parametrizing the AdS$_2$$\times$S$^2$ solution, corresponds to the ground state of the system with a given amount of flux. To find the symplectic form, we focus on the perturbations of this solution, which is equivalent to considering the small perturbations of the two harmonic functions $H_1$ and $H_2$. The profile can vibrate in $(x_1,\;x_2,\;y)$, but in this paper we only consider its fluctuations in the $(x_1,\;x_2)$ plane and restrict vector $\b$ to be orthogonal to this plane as (\ref{H}). We can express $\delta f$ $(f=\sqrt{f_1^2+f_2^2}$) in Fourier series:
\bea
\delta f=\sum_{|n|>1}L a_n e^{in\phi},~~~a^\ast_n=a_{-n}.\label{dx}
\eea
From now on we will set $L=1$, and this scale will be restored in the
 end [see Eqs. (\ref{BPbraket})-(\ref{qe})].
In this paper we need only the first order fluctuations of the profile since only such perturbations enter the expression of symplectic current~(\ref{current}).
In the first order the fluctuations with $n=\pm1$ describe the circular profile with shifted origin, so they still correspond to the ground state. Thus we focus on $|n|>1$.

The first order perturbations of the two harmonic functions $H_1$, $H_2$, and vector field $V$ caused by the fluctuations~(\ref{dx}) are given by
\bea\label{VandH}
\delta H_1&=&\sum_{|n|>1}\frac{a_ne^{in(\tilde{\phi}-\tilde{t})}}{2 R^6}\left[\frac{s_\theta}{\sqrt{\rho ^2+1}}\right]^{|n|}
\rho\left[\rho ^2-3 c^2_\theta \right] \left[|n| R^2+\rho ^2-c^2_\theta +2\right],
\nn\\
\delta H_2&=&\sum_{|n|>1} \frac{a_ne^{in(\tilde{\phi}-\tilde{t})}}{2 R^6}\left[\frac{s_\theta}{\sqrt{\rho ^2+1}}\right]^{|n|}
c_\theta\left[3 \rho ^2-c^2_\theta \right] \left[|n| R^2+\rho ^2-c^2_\theta +2\right],\nn\\
\delta V_r
&=&-\sum_{|n|>1}\frac{i n a_n e^{in(\tilde{\phi}-\tilde{t})} }{(\rho ^2+1) R^2}\left[\frac{s_\theta}{\sqrt{\rho ^2+1}}\right]^{\left| n\right|-1 },\\
\delta V_\phi
&=&\sum_{|n|>1}  \frac{a_ne^{in(\tilde{\phi}-\tilde{t})}}{R^6}{\left[\frac{s_\theta}{\sqrt{\rho ^2+1}}\right]^{\left| n\right| }}\left[|n|R^2\left(c_\theta^2-\rho^2+2\rho^2c_\theta^2\right)+2(\rho ^2+1)s^2_\theta(c_\theta^2-\rho ^2)\right],\nn
\eea
where $R^2=(\rho^2+\cos^2\theta)$, and $s_\theta\equiv\sin\theta,\;c_\theta\equiv\cos\theta$.  The details of the calculation can be seen in Appendix A. Although perturbations (\ref{VandH}) appear singular at the location of the profile ($R=0$), we will now show that they give rise to regular metric perturbations after an appropriate gauge transformation.

The metric perturbations are obtained by plugging the expressions~(\ref{VandH}) into~(\ref{GS}). In order to make the field perturbations regular, we perform a linear gauge transformation (\ref{gauge}) with
\bea
\xi=\sum_{|n|>1} a_n e^{in(\tilde{\phi}-\tilde{t})}\left[\frac{s_\theta}{\sqrt{\rho ^2+1}}\right]^{|n|}\left[-\frac{i|n| }{n}d\tilde t+\frac{ \rho  s ^2_\theta  }{\left(\rho ^2+1\right) R^2}d\rho+\frac{ s_{ \theta }c_\theta  }{ R^2}d\theta\right],
\eea
This leads to the final form of the metric perturbation:
\bea
h_{\mu\nu}
&=&\sum_{|n|>1}\bigg(-{2|n|(|n|-1)\/|n|+1}s_n Y_n g_{\mu\nu}+{4\/|n|+1}Y_n\nabla_{(\mu}\nabla_{\nu)}s_n\bigg),\label{mp}\nn\\
h_{\alpha\beta}&=&\sum_{|n|>1}2|n|s_nY_ng_{\alpha\beta},
\eea
where $\mu$, and $\nu$ run over the AdS$_2$ directions, $\alpha$, and $\beta$ run over the S$^2$ directions, and $s_n$ and $Y_n$ are
\bea
s_n&=& a_n e^{-in\tilde{t}}{|n|+1\/2|n|(1+\rho^2)^{|n|/2}},\nn\\
Y_n&=&e^{in\tilde{\phi}}\sin^{|n|}\theta.\label{snyn}
\eea
They are the spherical harmonics of AdS$_2$ and S$^2$, respectively,
\bea
\nabla^2_{\text{S}^2}Y_n=-|n|(|n|+1)Y_n,~~~
\nabla^2_{\text{AdS}_2}s_n=|n|(|n|-1)s_n.
\eea
The symplectic current of gravity $J^t_G$ is obtained by substituting~(\ref{mp}) into~(\ref{jg}).

The second part of the symplectic current, $J^l_F$, is constructed from the four-form potential according to~(\ref{jf}). For the solutions~(\ref{GS}), we need the fluctuations of the two-forms $F$ and $\tilde{F}$ in~(\ref{GS}) and their potentials. These fluctuations can be calculated by substituting~(\ref{dx}) or~(\ref{VandH}) into~(\ref{GS}),~(\ref{HF}) and~(\ref{H}), but this path involves tedious algebraic manipulations.

Fortunately, there is an alternative option given by~\cite{Grant,kim,lee}, which relates fluctuations of $F_5$ with the metric perturbations~(\ref{mp}). Following the procedure outlined in~\cite{Grant,kim,lee}, we can compute the one-form $B$ and $\tilde B$ through the relations
\bea
\delta a_{\alpha}&=&{1\/4}\epsilon_{\beta\alpha}s_n\nabla^{\beta}Y_n,\nn\\
\delta a_{\mu}&=&-{1\/4}\epsilon_{\nu\mu}Y_n\nabla^{\nu}s_n.\label{da}
\eea
Here $\mu$, $\nu$ run over the AdS$_2$ directions, $\alpha$, and $\beta$ run over the S$^2$ directions, and  $\epsilon_{\beta\alpha}$ and $\epsilon_{\mu\nu}$ are the volume forms on S$^2$ and AdS$_2$, respectively.
Plugging~(\ref{snyn}) into above relations~(\ref{da}), we immediately get
\bea
\delta B&=&-\sum_{|n|>1} {1\/8}a_n e^{in(\tilde{\phi}-\tilde{t})}\left[\frac{\sin \theta }{\sqrt{\rho ^2+1}}\right]^{|n|}(|n|+1)\left[\rho d\tilde{t}+\frac{i n  }{|n|(1+\rho^2)}d\rho\right],\nn\\
\delta \tilde{B}&=&-\sum_{|n|>1} {1\/8}a_n e^{in(\tilde{\phi}-\tilde{t})}\left[\frac{\sin \theta }{\sqrt{\rho ^2+1}}\right]^{|n|}(|n|+1)\left[\frac{i n }{ |n| \sin\theta} d\theta-\cos\theta d\tilde{\phi}\right],\label{dA}
\eea
and
\bea\label{df}
\delta F&=&-\sum_{|n|>1} {1\/8}a_n e^{in(\tilde{\phi}-\tilde{t})}\left[\frac{\sin \theta }{\sqrt{\rho ^2+1}}\right]^{|n|}(|n|+1)\bigg[(|n|-1)d\tilde{t}\wedge d\rho-|n|\cot\theta d\tilde{t}\wedge d\theta\nn\\
&&~~-in\rho d\tilde{t}\wedge d\tilde{\phi}-{in\cot\theta \/\rho^2+1} d\rho\wedge d\theta+{|n|\/\rho^2+1}d\rho\wedge d\tilde{\phi} \bigg],\nn\\
\delta\tilde{F}&=&-\sum_{|n|>1} {1\/8}a_n e^{in(\tilde{\phi}-\tilde{t})}\left[\frac{\sin \theta }{\sqrt{\rho ^2+1}}\right]^{|n|}\left(|n|+1\right)\bigg[(|n|+1)\sin\theta d\theta\wedge d\phi+{|n|\csc\theta}d\tilde t\wedge d\theta\nn\\
&&~~+in\cos\theta d\tilde t \wedge d \tilde \phi-{in\rho\csc\theta\/\rho^2+1}d\rho\wedge d\theta+{|n|\rho\cos\theta\/\rho^2+1}d\rho\wedge d\tilde \phi\bigg].
\eea

It is convenient to pick the hypersurface $\Sigma=\{\tilde{t}=\text{const}\}$ in the symplectic form~(\ref{SF}), since the solutions~(\ref{GS}) are independent of $\tilde t$. This leads to the gravity and to five-form contributions to the symplectic form:
\bea
\int_{\tilde{t}=const}J^{\tilde t}_G&=&-\sum_{|n|>1}{2\pi^2iV_6|n|(|n|+1)(n^2-3|n|-2)\/n(2|n|+1)}a_n\wedge a_{-n},\label{JG}\\
\int_{\tilde{t}=const}J^{\tilde t}_F&=&-\sum_{|n|>1} {2\pi^2iV_6|n|(|n|+1)^3\/n(2|n|+1)}a_n\wedge a_{-n},\label{JF}
\eea
where $V_6$ is the volume of T$^6$. In Eqs.~(\ref{JG}) and~(\ref{JF}) the Fourier coefficients
$a_n$ should be interpreted as one-forms on the phase space of solutions~(\ref{GS}). The derivation  of~(\ref{JG}) and~(\ref{JF}) can be found  in Appendix B. Adding the individual contributions, we get the symplectic form as
\bea
\Omega
&=&-{2\pi^2iV_6\/\kappa^2_{10}}\sum_{|n|>1}\text{sign}\;n\;(n^2-1)\;a_n\wedge a_{-n}.\label{FSF}
\eea

As we mentioned in Sec. III, once the symplectic form is found, the Poisson brackets can be extracted from it following the procedure summarized by~(\ref{lsf}) and~(\ref{PB}). Restoring the radius $L$, we obtain
\beq\label{BPbraket}
\{a_n,\;a_m\}={L^2\kappa^2_{10}\/4\pi^2V_6}\;{i\;\text{sign}\;n\/n^2-1}\delta_{m+n}.
\eeq
The commutators are immediately  obtained in the usual way by using the relation that $[\;,\;]=i\{\;,\;\}$
\beq
[a_m,\;a_n]={L^2\kappa^2_{10}\/4\pi^2V_6}\;{\;\text{sign}\;n\/n^2-1}\delta_{m+n}.\label{com}
\eeq
The microstate geometries in supergravity correspond to different profiles $\f(v)$. Although classically there is a continuum of such curves, the commutators~(\ref{com}) lead to expansions into discrete quantum oscillators,
\bea\label{qe}
&&f(\phi)=f_0(\phi)+\lambda\sum^\infty_{n=2}{1\/\sqrt{n^2-1}}\left(c_ne^{-in\phi}+c^\dagger_ne^{in\phi}\right),\nn \\
&&[c_n,\;c^\dagger_m]=\delta_{mn},
\eea
where
\bea
\lambda={L\kappa_{10}\/2\pi\sqrt{V_6}},~~~c_n=\left[{L^2\kappa^2_{10}\/4\pi^2V_6}\;{1\/n^2-1}\right]^{-1/2}a_{-n}.\nn
\eea
The Hilbert space of this theory is a bosonic Fock space equipped with annihilation operators $c_n$ and creation operators $c^\dagger_n$.

The counterparts of the commutators~(\ref{qe}) have been encountered in
 \cite{SR,Grant,LM}, and these quantization conditions derived in
 supergravity turned out to be in a perfect agreement with properties
 of supersymmetric states in $N$ = 4 super Yang-Mills on S$^3$$\times$$\mathbb{R}$~\cite{Cor,Berenstein:2004kk}
 and in the two-dimensional orbifold CFT~\cite{OM, LM02}. We expect that a
 better understanding for the boundary theory in the AdS$_2$ case would
 lead to a similar agreement for our relations~(\ref{qe}).

\section{Discussion}

In this paper, we focused on regular horizon-free supersymmetric solutions describing normalizable excitations of AdS$_2$$\times$S$^2$, which are parametrized by a complex harmonic function $H$ in $\mathbf{R}^3$ of possessing the sources located in closed curves. From the gravity side, we have generalized the method used in~\cite{SR,Grant,LM} to quantize the moduli space of these solutions directly. The basic idea of this approach is that commutation relations can be encoded from the symplectic form. Starting with the profile fluctuations, we obtain the perturbations of fields, graviton and five-form, involved in Type IIB SUGRA action corresponding to solutions~(\ref{GS}). Finally these perturbations are used to find the symplectic form and to quantize the profiles describing microscopic states. As in the AdS$_3$ and AdS$_5$ cases discussed in \cite{SR,Grant,LM}, this quantization is expected to agree with results in the dual theory on the boundary. However, in the present situation where the boundary theory is poorly understood, our results can be viewed as predictions rather than consistency checks as in \cite{SR,Grant,LM}.

\begin{acknowledgments}

I am very grateful to my advisor Oleg Lunin for lots of discussions and for his patient guidance over this work. I also thank Jia Tian for useful discussions.

\end{acknowledgments}

\begin{appendix}

\section{Shifts in harmonic functions and vector fields}

\subsection{Evaluation of $\delta H_1$ and $\delta H_2$}

The vectors $\r,\;\f\;,\b$ in~(\ref{H}) can be written in the cylindrical coordinates as
\bea
\r&=&(r\cos\phi,\;r\sin\phi,\;y),\nn\\
\f&=&(f\cos\phi_1,\;f\sin\phi_1,\;0),\nn\\
\b&=&(b\cos\phi_2,\;b\sin\phi_2,\;ib_3),\label{cyl}
\eea
where $r,f,b,b_3$ are real. In this paper we only consider the planar profiles, thus taking $f_3=0$.
Applying (\ref{cyl}) and picking $v=f\phi_1$, Eq.~(\ref{H}) becomes
\bea
H&=&{1\/2\pi}\int d\phi_1{f\sigma\/{r^2+f^2+y^2-2fr\cos(\phi_1-\phi)}}\nn\\
&&~~\times\sqrt{r^2+f^2+y^2-2fr \cos(\phi_1-\phi)+br \cos(\phi_2-\phi)-bf\cos(\phi_2-\phi_1)+iyb_3}\nn\\
&=&{1\/2\pi}\int  C \sqrt{D+iyb_3}\;\;d\phi_1,
\eea
where
\bea
C&=&{f\sigma\/{r^2+f^2+y^2-2fr\cos(\phi_1-\phi)}},\nn\\
D&=&r^2+f^2+y^2-2fr \cos(\phi_1-\phi)+br \cos(\phi_2-\phi)-bf\cos(\phi_2-\phi_1).\nn
\eea
Therefore, we can express $H_1, H_2$ as
\bea
H_1&=&{1\/2\pi}\int^{2\pi}_0 C \left[D^2+(yb_3)^2\right]^{{1\/4}} \cos\bigg[{1\/2}\tan^{-1}\left({{yb_3}\/D}\right)\bigg]\;\;d\phi_1,\label{H11}\\
H_2&=&{1\/2\pi}\int^{2\pi}_0 C \left[D^2+(yb_3)^2\right]^{{1\/4}} \sin\bigg[{1\/2}\tan^{-1}\left({{yb_3}\/D}\right)\bigg]\;\;d\phi_1.\label{H22}
\eea

The profile corresponding to AdS$_2$$\times$S$^2$ is given by~(\ref{adsf}), and the corresponding vector $\b$ is given as $\b=2L\left(\cos{v\/L},\;\sin{v\/L},\;i\right)$, where $L$ is the radius of the AdS and the sphere. We set $L=1$ for convenience. Now we consider perturbations~(\ref{dx}) on this profile, i.e., now $\f=(1+\delta f)(\cos\phi_1,\;\sin\phi_1,\;0)$. Vector $\b$ has to change correspondingly due to the constraint $\b\cdot\dot{\f}=0$,
\bea
\b=\b_0+\delta\b,
\eea
where
\bea
\b_0&=&2L\left(\cos{v\/L},\;L\sin{v\/L},\;i\right)=2(\cos\phi_1,\;\sin\phi_1,\;i)=2(\cos\phi_2,\;\sin\phi_2,\;i),\nn\\
\delta\b&=&2\left(-\sin\phi_2,\;\cos\phi_2,\;0\right)\delta\phi_2=2\left(-\sin\phi_1,\;\cos\phi_1,\;0\right)\delta\phi_2.
\eea
Here we keep $b=2$ unchanged, while $\phi_2\rightarrow\phi_2+\delta\phi_2$, which can guarantee the two constraints $\b\cdot\dot{\f}=0,\;\b\cdot\b=0$.
Using constraint
 \bea
 0=\b\cdot\dot{\f}=\delta\dot{f}+(1+\delta f)\delta\phi_2\approx\delta\dot{f}+\delta\phi_2,
 \eea
we find $\delta\phi_2=-\delta\dot{f}=-\sum_n in a_n e^{in\phi_1}$.

 Thus we can express the first order perturbations of harmonic function $H_1$ and $H_2$ as
\bea
\delta H_1=\left[{\partial H_1\/\partial f}\delta f+{\partial H_1\/\partial \phi_2}(-\delta \dot{f})\right]_{\sigma=1,f=1,b=b_3=2,\phi_1=\phi_2},\nn\\
\delta H_2=\left[{\partial H_2\/\partial f}\delta f+{\partial H_2\/\partial \phi_2}(-\delta \dot{f})\right]_{\sigma=1,f=1,b=b_3=2,\phi_1=\phi_2}.
\eea
Let $\phi_1-\phi=\alpha$, then $\phi_1=\phi+\alpha$, and we obtain
\bea
&&\left[{\partial H_1\/\partial f}\delta f\right]_{\sigma=1,f=1,b=b_3=2,\phi_1=\phi_2}\nn\\
&&=\int d\alpha {{\sum_{|n|>1}a_n} e^{in(\phi+\alpha)}\/2\pi\left[\left(r^2+y^2-1\right)^2+4 y^2\right]^{3/4} \left(r^2-2 r \cos \alpha+y^2+1\right)^2}\nn\\
&&~~\times\bigg\{(r^2+y^2-1)\big[r^4+r \left(r \cos 2 \alpha-\left(r^2+y^2+1\right) \cos \alpha\right)\nn\\
&&~~~~~~~+r^2 \left(2 y^2-1\right)+\left(y^2+1\right)^2\big]\cos \left[\frac{1}{2} \tan ^{-1}\left(\frac{2 y}{r^2+y^2-1}\right)\right]\nn\\
&&~~~~~~~-2 r y \cos \alpha \sin \left[\frac{1}{2} \tan ^{-1}\left(\frac{2 y}{r^2+y^2-1}\right)\right] \left(r^2-2 r \cos \alpha+y^2+1\right)\bigg\}.
\eea
Working out this integral over $\alpha$ by the residue theorem, we find first part of $\delta H_1$. Similarly, we can figure out another part, and finally we obtain, after transferring to global AdS coordinates, the perturbations of $H_1$. Repeating the same procedure, we will find the first  order shifts of $\delta H_2$.

\subsection{Evaluation of $\delta V_r$ and $\delta V_\phi$}

To find  $\delta V$ corresponding to $\delta H$, we begin with equation
\beq
dV=-2\star_3[H_2dH_1-H_1dH_2].\label{dv}
\eeq
The  exterior derivative of vector field $V$, in the gauge $V_y$=0, can be written as:
\bea
dV&=&{\partial V_r\/\partial y}dy\wedge dr+{\partial V_\phi\/\partial y}dy\wedge d\phi+\left({\partial V_\phi\/\partial r}-{\partial V_r\/\partial \phi}\right)dr\wedge d\phi.
\eea
Through the comparison of the left and right hand sides of (\ref{dv}), we get the differential equations
\bea
{\partial V_r\/\partial y}&=&{2\/r}\left(H_1{\partial H_2\/\partial \phi}-H_2{\partial H_1\/\partial \phi}\right),\nn\\
{\partial V_\phi\/\partial y}&=&-{2r}\left(H_1{\partial H_2\/\partial r}-H_2{\partial H_1\/\partial r}\right),\nn\\
{\partial V_\phi\/\partial r}-{\partial V_r\/\partial \phi}&=&{2r}\left(H_1{\partial H_2\/\partial y}-H_2{\partial H_1\/\partial y}\right).\label{dequ}
\eea

Considering the first order shifts of $V_r$ and $V_\phi$, we get
\bea
{\partial( \delta V_r)\/\partial y}&=&{2\/r}\bigg[\left({\partial H_1\/\partial f}\delta f-{\partial H_1\/\partial\phi_2}\delta\dot{f}\right){\partial H_2\/\partial\phi}
+H_1{\partial\/\partial \phi}\left({\partial H_2\/\partial f}\delta f-{\partial H_2\/\partial\phi_2}\delta\dot{f}\right)\nn\\
&&~~~-\left({\partial H_2\/\partial f}\delta f-{\partial H_2\/\partial\phi_2}\delta\dot{f}\right){\partial H_1\/\partial\phi}
-H_2{\partial\/\partial \phi}\left({\partial H_1\/\partial f}\delta f-{\partial H_1\/\partial\phi_2}\delta\dot{f}\right)\bigg]_{f=1}.
\eea
Similarly, we have
\bea
{\partial( \delta V_\phi)\/\partial y}&=&-{2r}\bigg[\left({\partial H_1\/\partial f}\delta f-{\partial H_1\/\partial\phi_2}\delta\dot{f}\right){\partial H_2\/\partial r}
+H_1{\partial\/\partial r}\left({\partial H_2\/\partial f}\delta f-{\partial H_2\/\partial\phi_2}\delta\dot{f}\right)\nn\\
&&~~~-\left({\partial H_2\/\partial f}\delta f-{\partial H_2\/\partial\phi_2}\delta\dot{f}\right){\partial H_1\/\partial r}
-H_2{\partial\/\partial r}\left({\partial H_1\/\partial f}\delta f-{\partial H_1\/\partial\phi_2}\delta\dot{f}\right)\bigg]_{f=1}.\label{dvy}
\eea
Again, letting  $\phi_1-\phi=\alpha$, we obtain
\bea
{\partial (\delta V_r)\/\partial y}&=&\sum_{|n|>1}{ a_n e^{i n \phi}\/2\pi}\int^{2\pi}_0\frac{4  n y  (n \sin \alpha +i \cos \alpha )e^{i n \alpha }d\alpha}{\left[r^4+2 r^2 \left(y^2-1\right)+\left(y^2+1\right)^2\right] \left(r^2-2 r \cos \alpha +y^2+1\right)}.\label{dv1}\nn\\
\eea
Completing this integral over $\alpha$, we find $\delta V_r$ in the global AdS coordinates.
Proceeding with~(\ref{dvy}), we obtain
\bea
{\partial (\delta V_\phi)\/\partial y}&=&
\sum_{|n|>1}{a_n e^{i n \phi}\/2 \pi}\int^{2\pi}_0  \frac{4 r y \left[ \cos \alpha  \left(r^2-y^2-1\right)+2  r \left(r^2+y^2-1\right)-in \sin \alpha  \left(r^2-y^2-1\right)\right]}{\left[r^4+2 r^2 \left(y^2-1\right)+\left(y^2+1\right)^2\right] \left(r^2-2 r \cos \alpha +y^2+1\right)^2}\nn\\
&&~~~~\times e^{in\alpha}d\alpha.\label{dv2}
\eea
Figuring out this integral and changing the coordinates as~(\ref{adsc}), we immediately find $\delta V_\phi$. It is easy to check that $\delta V_r$ and $\delta V_\phi$ obtained from~(\ref{dv1}) and~(\ref{dv2}) are satisfied by the third equation in~(\ref{dequ}).

\section{Symplectic currents}

\subsection{Evaluation of symplectic current $J^t_G$}

From metric perturbation~(\ref{mp}), it is straightforward to find  the components of $\delta[g^{mn}\sqrt{-g}\;]$:
\bea
\delta[g^{\tilde{t}\tilde{t}}\sqrt{-g}]&=&-\sum_{|n|>1} 2a_n e^{in(\tilde{\phi}-\tilde{t})}{(\left|n\right|+1)\rho^2\sin\theta\mathcal{S}_n\/(1+\rho^2)^2},\nn\\
\delta[g^{\tilde{t}\rho}\sqrt{-g}]&=&\sum_{|n|>1} 2a_n e^{in(\tilde{\phi}-\tilde{t})}{i (n^2+|n|)\rho\sin\theta\mathcal{S}_n\/n(1+\rho^2)},\nn\\
\delta[g^{\rho\rho}\sqrt{-g}]&=&\sum_{|n|>1} 2a_n e^{in(\tilde{\phi}-\tilde{t})}{(|n|+1)\sin\theta}\mathcal{S}_n,\label{dgg}
\eea
where \beq\mathcal{S}_n=\left(\frac{\sin \theta }{\sqrt{\rho ^2+1}}\right)^{|n|}.\nn\eeq
The shifts of connections $\delta\Gamma^{\tilde{t}}_{mp}$ that we are interested in are
\bea
\delta\Gamma^{\tilde{t}}_{\tilde{t}\tilde{t}}&=&\sum_{|n|>1} a_n e^{in(\tilde{\phi}-\tilde{t})}{i\big[n^2\left( |n| \left(\rho ^2-1\right)+5 \rho ^2-1\right)+4 |n| \rho ^2\big]\mathcal{S}_n\/2n(1+\rho^2)},\nn\\
\delta\Gamma^{\tilde{t}}_{\tilde{t}\rho}&=&\sum_{|n|>1} a_n e^{in(\tilde{\phi}-\tilde{t})}{\rho (|n|+1)  \left[|n| \left(\rho ^2-1\right)-4\right]\mathcal{S}_n\/2(1+\rho^2)^2},\nn\\
\delta\Gamma^{\tilde{t}}_{\rho\rho}&=&\sum_{|n|>1} a_n e^{in(\tilde{\phi}-\tilde{t})}{i\big[n^2 \left(3 |n| \rho ^2+|n|-\rho ^2-3\right)+4 |n| \left(|n| \rho ^2-1\right)\big]\mathcal{S}_n\/2n(1+\rho^2)^3}.\nn\\ \label{dgamma}
\eea
Some components of $\delta\Gamma^m_{mp}$ are:
\bea
\delta\Gamma^m_{m\tilde{t}}&=&-\sum_{|n|>1}a_n e^{in(\tilde{\phi}-\tilde{t})}in(|n|+1)\mathcal{S}_n,\nn\\
\delta\Gamma^m_{m\rho}&=&-\sum_{|n|>1}a_n e^{in(\tilde{\phi}-\tilde{t})}{|n|(|n|+1)\rho\mathcal{S}_n\/1+\rho^2}.\label{dg2}
\eea
With the preparation~(\ref{dgg}),~(\ref{dgamma})and~(\ref{dg2}), we can easily find
\bea
-\int d\tilde{\phi}\;\delta\Gamma^{\tilde{t}}_{mp}\wedge\delta[\sqrt{-g}g^{mp}]&=&{2\pi i|n|(1+|n|)^2\big[|n| \left(3 \rho ^4-6 \rho ^2-1\right)+4 \left(\rho ^2-1\right)^2]\sin\theta\mathcal{S}^2_n\/n(1+\rho^2)^3}\nn\\
&&~~~~~\times(a_n\wedge a_{-n}),\\
\int d\tilde{\phi}\;\delta\Gamma^p_{mp}\wedge\delta[\sqrt{-g}g^{m\tilde{t}}]&=&{8\pi i(|n|+n^2)^2\rho^2\sin\theta\mathcal{S}^2_n\/n(1+\rho^2)^2}(a_n\wedge a_{-n})\;.
\eea
Integrate the remaining integrals, and add them up we then find the symplectic form~(\ref{JG}) from gravity current.

\subsection{Evaluation of symplectic current $J^t_F$}

Using relations~(\ref{da}), one can easily find
\bea
\delta B_{\tilde{t}}&=&-\sum_{|n|>1} {1\/8}a_n e^{in(\tilde{\phi}-\tilde{t})}(1+|n|)\rho\mathcal{S}_n,\nn\\
\delta B_\rho&=&-\sum_{|n|>1} a_n e^{in(\tilde{\phi}-\tilde{t})}\frac{i n (|n|+1)\mathcal{S}_n  }{8|n|(1+\rho^2)},\nn\\
\delta \tilde B_\theta&=&-\sum_{|n|>1} a_n e^{in(\tilde{\phi}-\tilde{t})}\frac{i n (|n|+1)\mathcal{S}_n}{8 |n| \sin\theta},\nn\\
\delta \tilde B_{\tilde{\phi}}&=&\sum_{|n|>1} {1\/8}a_n e^{in(\tilde{\phi}-\tilde{t})}(|n|+1)\cos\theta\mathcal{S}_n.
\eea

From these one-form perturbations we obtain
\bea
\delta F_{\tilde{t}\rho}&=&-\sum_{|n|>1} {1\/8}a_n e^{in(\tilde{\phi}-\tilde{t})}(n^2-1)\mathcal{S}_n,\nn\\
\delta \tilde F_{\tilde{t}\theta}&=&-\sum_{|n|>1} {1\/8}a_n e^{in(\tilde{\phi}-\tilde{t})}|n|(|n|+1)\csc\theta\mathcal{S}_n,\nn\\
\delta \tilde F_{\tilde{t}\tilde{\phi}}&=&-\sum_{|n|>1} {1\/8}a_n e^{in(\tilde{\phi}-\tilde{t})}in (|n|+1)\cos\theta\mathcal{S}_n.
\eea

Then we obtain
\bea
\delta\big[\sqrt{-g}F^{\tilde{t}\rho}\big]&=&\sum_{|n|>1} {1\/8}a_n e^{in(\tilde{\phi}-\tilde{t})} (|n|+1)^2\sin\theta\mathcal{S}_n,\nn\\
\delta\big[\sqrt{-g}\tilde F^{\tilde{t}\theta}\big]&=&\sum_{|n|>1} {1\/8}a_n e^{in(\tilde{\phi}-\tilde{t})}{|n|(|n|+1)\mathcal{S}_n\/1+\rho^2},\nn\\
\delta\big[\sqrt{-g}\tilde F^{\tilde{t}\tilde{\phi}}\big]&=&\sum_{|n|>1} {1\/8}a_n e^{in(\tilde{\phi}-\tilde{t})}{in(|n|+1)\cot\theta\mathcal{S}_n\/1+\rho^2}.
\eea

Therefore we find
\bea
\int d\tilde{\phi}\;8\;\delta A_{|i_1i_2i_3i_4|}\wedge\delta\big[\sqrt{-g}F^{t|i_1i_2i_3i_4|}\;\big]&=&-\sum_{|n|>1}{\pi i|n|(|n|+1)^2\left(2|n|+\sin^2\theta\right)\mathcal{S}^2_n\/n(1+\rho^2)\sin\theta}\nn\\
&&~~~\times  a_n\wedge a_{-n}.
\eea
 Working out the remaining integrals, we then find the symplectic form~(\ref{JF}) from five-form current.

\end{appendix}

\end{document}